\documentclass[twocolumn,showpacs,preprintnumbers,amsmath,amssymb,prl]{revtex4-1}

\usepackage{graphicx}
\usepackage{dcolumn}
\usepackage{bm}

\usepackage[latin1]{inputenc}

\begin{document}

\title{Criticality of an isotropic-to-smectic transition \\ induced by anisotropic quenched disorder}

\author{Gilbert Chahine$^{1}$}
\altaffiliation[Also at ]{Laboratoire Léon Brillouin, CEA Saclay.}
\author{Andriy V. Kityk$^{2}$}%
\author{Klaus Knorr$^{3}$}%
\author{Ronan Lefort$^{1}$}
\author{Mohammed Guendouz$^{4}$}
\author{Denis Morineau$^{1}$}
\author{Patrick Huber$^{3}$}%
\affiliation{%
$^{1}$Institut de Physique de Rennes, UMR 6251, Université de Rennes 1, 35042 Rennes, France
}%
\affiliation{%
$^{2}$Institute for Computer Science, Czestochowa University of Technology, 42200 Czestochowa, Poland
}%
\affiliation{%
$^{3}$Faculty of Physics and Mechatronics Engineering, Saarland University, 66041 Saarbrücken, Germany
}%

\affiliation{%
$^{4}$Laboratoire FOTON, UMR 6082, Université de Rennes 1, 22302 Lannion, France
}%

\date{\today}

\begin{abstract}
We report combined optical birefringence and neutron scattering measurements on the liquid crystal 12CB nanoconfined in mesoporous silicon layers. This liquid crystal exhibits strong nematic-smectic coupling responsible for a discontinuous isotropic-to-smectic phase transition in the bulk state. Confined in porous silicon, 12CB is subjected to strong anisotropic quenched disorder: a short-ranged smectic state evolves out of a paranematic phase. This transformation appears continuous, losing its bulk first order character. This contrasts with previously reported observations on liquid crystals under isotropic quenched disorder. In the low temperature phase, both orientational and translational order parameters obey the same power-law.
\end{abstract}

\pacs{64.70.Nd, 61.30.Eb}
                           
\maketitle

Since its introduction in 1975 by Imry and Ma \cite{PhysRevLett.35.1399}, the influence of random fields on phase transitions has been one of the most debated topics in condensed matter physics. Various realizations of this phenomenology could be achieved by confining liquid crystals (LCs) in geometrically disordered porous materials or gels \cite{Bellini2001, *guegan:011707, *Guegan2008, *lefort:040701, Lehenypre, Liang-8CB-aerosil-PRE-2007, *Iannacchine-8CB-PRE-2009, *Iannacchine-8CB-JPCB-2009}. In principle, the geometric restriction introduces two forms of disorder in LCs: random orientational fields that couple to the nematic director and random positional fields that couple to the smectic order. Studies on the influence of these couplings while varying the magnitude of disorder and its isotropic or anisotropic character has largely contributed to recent progress in the understanding of quenched disorder (QD) effects on phase transitions in LCs \cite{PhysRevLett.79.4214,*PhysRevB.60.206,Bellini2001, Terentjev2009}. Owing to their generic character, these findings are also relevant to many other systems, most prominently for supraconductors or for superfluids \cite{Gennes1976(2ndedition)}. 

Both the second order nematic-smectic A (N-SmA) and normal-superconducting transitions can be mapped onto each other and, in principle, fall in the universality class of the 3D~$XY$ model  \cite{Gennes1976(2ndedition), PhysRevE.49.2964}, since both can be described by a complex order parameter representing the amplitude and phase of a sinusoidal-varying smectic mass density wave or a macroscopic wave function, respectively. However, as was pointed out first by de Gennes the coupling between nematic (Q) and smectic ($\eta$) order parameters (OPs) actually takes the N-SmA transition away from that class. Conversely, increasing the strength of isotropic QD can gradually shift the character of the N-SmA transition from tricritical back to 3D~$XY$ universality. This was verified by studies on \textit{n}-octyl-cyanobiphenyl (8CB) in aerogels or loaded with random dispersions of aerosil particles \cite{Lehenypre,Liang-8CB-aerosil-PRE-2007,*Iannacchine-8CB-PRE-2009,*Iannacchine-8CB-JPCB-2009}. For these systems, the influence of strong isotropic QD in the weak $Q$-$\eta$ coupling limit was addressed. For anisotropic QD, this striking effect is already visible in the limit of weak disorder strength in the case of 8CB. Following the argumentation of Garland and Iannachione \cite{Liang-8CB-aerosil-PRE-2007,*Iannacchine-8CB-PRE-2009,*Iannacchine-8CB-JPCB-2009}, this suggests that the $Q$-$\eta$ coupling is reduced by isotropic QD, and is even entirely turned off by anisotropic QD. 

An interesting test of this hypothesis was made by Ramazoglu \textit{et. al.} \cite{10CB-aerosil-PhysRevE.2004} on the 10CB/aerosil composite. In the bulk, this LC shows a strong $Q$-$\eta$ coupling, which leads to a direct and discontinuous isotropic (I) to SmA transition. Under isotropic QD, the transition remains first order but the temperature scaling associated to the growth of the order parameter tends towards tricritical behavior as the density of disorder is increased \cite{Bellini-10CB-JPCM-2003,*10CB-aerosil-PhysRevE.2004}. The emerging question is whether a direct I-SmA transition of a strongly $Q$-$\eta$ coupled LC could become continuous under strong \textit{anisotropic} quenched disorder.

Here we address this ultimate case of anisotropic QD effects on a LC with strong $Q$-$\eta$ coupling. We performed complementary optical birefringence and neutron diffraction experiments on \textit{n}-dodecyl-1,4-cyanobiphenyl (12CB), a rod-like mesogen, spatially confined in columnar porous silicon (pSi) or porous silica (pSiO$_2$) layers. These substrates are known to induce anisotropic QD \cite{*guegan:011707, *Guegan2008, *Wallacher2004, *Naumov2009}. Both nematic $Q$ and smectic $\eta$ OPs are measured simultaneously, which allows us to provide unique information both about the transition mechanism and the hypothetical reduction of the $Q$-$\eta$ coupling. The somewhat surprising experimental findings shall be compared with predictions of state-of-the-art theories of constrained smectic order.

\begin{figure}
\includegraphics[width=0.44\textwidth]{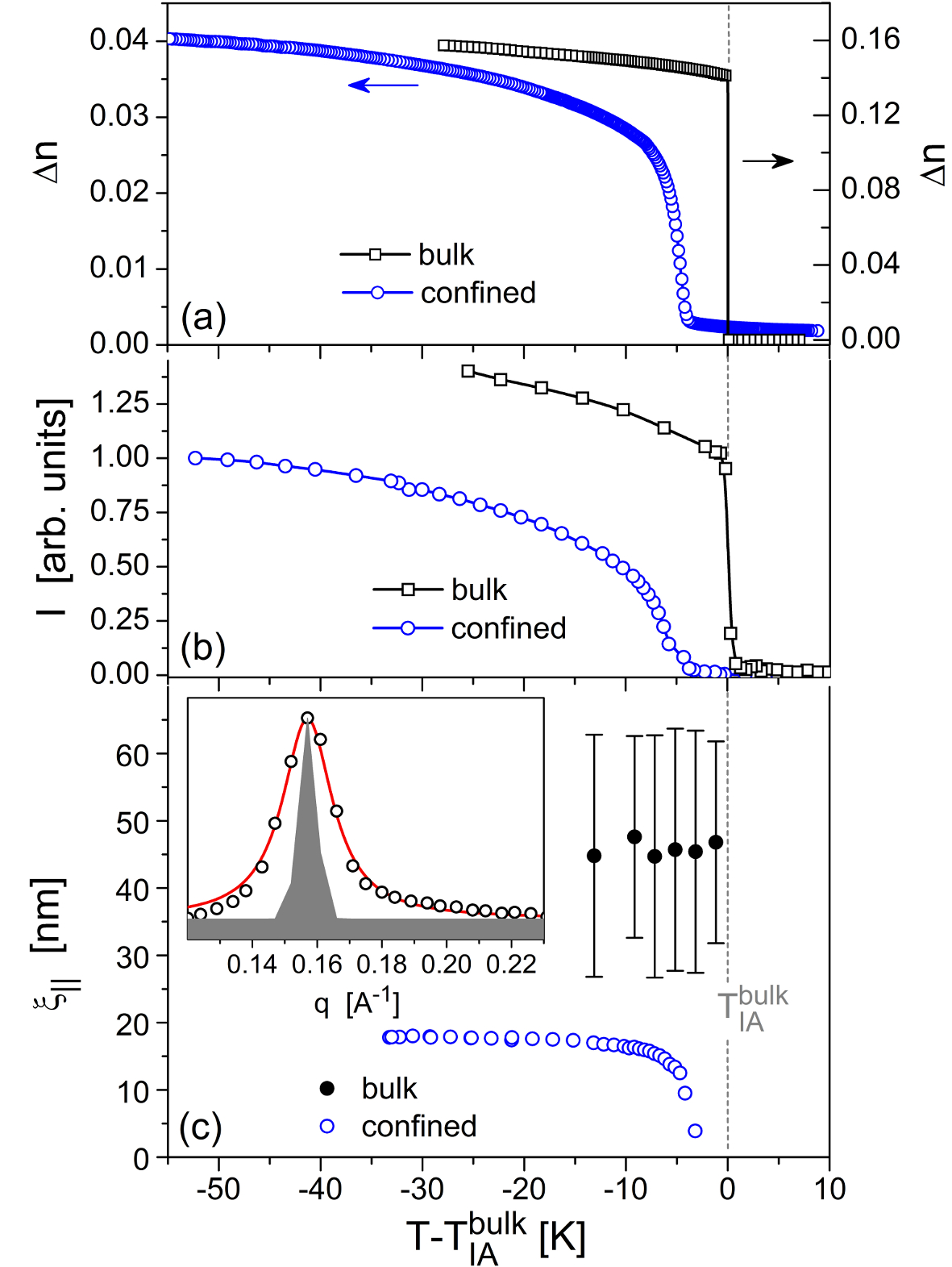} 
\caption{\label{fig:1}(Color online) (a) Birefringence and (b) integrated intensity of the smectic neutron diffraction peak of 12CB measured upon slow cooling, in the bulk (squares) and confined in porous SiO$_2$ or porous Si (circles). (c) Smectic correlation lengths of bulk 12CB (average isotropic $\xi_{iso}$, solid circles) and 12CB confined in pSi ($\xi_{\parallel}$, open circles). Inset: smectic peak shape of confined 12CB at T=326 K (experimental resolution in filled gray area, the solid line is the best fit according to eqn.\ref{eqn:Grin}, see text).}
\end{figure}

12CB was purchased from Synthon GmbH and used without further purification. This LC compound displays the following bulk phase sequence : I for T $>$ 331 K ; SmA for 305 K $<$ T $<$ 331 K and crystalline for T $<$ 305 K.
For neutron scattering experiments, pSi films were prepared by electrochemical etching of a crystalline, highly doped Si wafer \cite{fabripSi}. No special attempt was made to remove the native silicon oxide from the inner walls of the porous structure. For birefringence measurements, a monolithic silica film was prepared in a similar way, followed by thermal oxidation at 800$^\circ$C for 12~h \cite{gruener:064502,*kumar:024303}. All resulting films exhibit a structure permeated by an array of parallel-aligned, non-interconnected channels of 300 $\mu$m length and $\sim$ 10 nm diameter. Transmission electron micrographs of the channels \cite{gruener:064502, *kumar:024303} indicate sizeable 1.0 $\pm$  0.5 nm mean square deviations of their surfaces from an ideal cylindrical form. 
The films were completely filled with the LCs by spontaneous imbibition in the isotropic phase. Diffraction experiments were carried out on the two-axis spectrometer G6.1 at the Léon Brillouin Laboratory (CEA Saclay). 
The porous films were macroscopically oriented in grazing incidence scattering geometry in order to probe momentum transfers parallel to the pores axis. The smectic A phase of 12CB is characterized by a single observable Bragg peak at $q_0=$0.158 \AA$^{-1}$ (see inset of Fig.\ref{fig:1}c). Its integrated intensity is proportional to the square of the smectic order parameter, $I(T)\propto \vert\eta\vert^2$.
In agreement with measurements on 8CB \cite{Guegan2008}, this peak appears only for \textbf{q} vectors aligned with the pSi layer surface normal, indicating an homogeneous uniaxial macroscopic  ordering of the smectic stacking along the long axes of the pores \cite{PhysRevE.54.1610}.

Because of its optical transparency and straight pore geometry the pSiO$_2$ is particularly suitable for optical polarization studies. Accordingly, the orientational molecular ordering inside the pores can be precisely evaluated by optical birefringence measurements performed in transmission geometry \cite{kityk:187801}. To a good approximation, the optical birefringence in the nematic or smectic phase is proportional to $Q$ \cite{Haller1975}. 

Fig.\ref{fig:1} a and b display the $T$-dependency of the optical birefringence $\Delta n(T)$ and integrated intensity $I(T)$ of the neutron diffraction smectic peak measured for 12CB in the bulk state and under confinement. All measurements have been performed upon slow cooling with a rate of 0.05$^\circ$ C/min. The neutron data are displayed either normalized to the value at $T_{IA}^{bulk}$ (bulk), or to their maximum value (confined samples). As expected for a first order I-SmA transition, the two OPs of bulk 12CB simultaneously jump at $T_{IA}$, with no detectable pretransitional effects in the isotropic phase. The magnitude of the jump at the transition $\Delta n(T) \approx$ 0.141 well agrees with previous studies \cite{PhysRevE.64.010702}. Confinement obviously induces fundamental modifications of the $T$-behavior of the two OPs. Even for temperatures above $T_{IA}^{conf}$ there exists a weak residual birefringence, characteristic of nematic ordering along the pore axis. Interestingly, this paranematic order induced by confinement is not accompanied by a similar parasmectic order ($I(T>T_{IA}^{conf})=0$). More important, the first-order, discontinuous character of the bulk $I-A$ transition is replaced by a continuous evolution of both the orientational (nematic) and translational (smectic) ordering as a function of $T$. Both OPs indicate a transition onset temperature $T_{IA}^{conf}$ which is about 4~K below the bulk transition temperature.
At lower temperatures, in the confined SmA phase, the absolute magnitude of $\Delta n$ (after the normalization on porosity $P$) is roughly compatible with only 75\% of its magnitude in bulk 12CB. 

Interestingly, in the confined SmA phase, smectic order remains short-ranged, as indicated by a broad Bragg peak in Fig.\ref{fig:1}c. This observation can be attributed to the aforementioned susceptibility of the quasi-long range smectic order to QD effects introduced by the pore confinement \cite{Bellini2001, *guegan:011707,*Guegan2008,*lefort:040701, Lehenypre}. Accordingly, we managed to fit the shape of this diffraction peak by a structure factor characteristic of short-ranged smectic order \cite{PhysRevB.27.4503} while using a partial powder averaging procedure, described elsewhere \cite{guegan:011707,*lefort:040701}:
\begin{eqnarray}
\label{eqn:Grin}
I(q)=\dfrac{A_{therm}}{1+\xi_{\parallel}^2(q_{\parallel}-q_0)^2+\xi_{\perp}^2q_{\perp}^2(1+c\xi_{\perp}^2q_{\perp}^2)}\nonumber\\
+\dfrac{A_{disorder}}{[1+\xi_{\parallel}^2(q_{\parallel}-q_0)^2+\xi_{\perp}^2q_{\perp}^2(1+c\xi_{\perp}^2q_{\perp}^2)]^2}
\end{eqnarray}

In Eq. \ref{eqn:Grin}, the first term describes thermally-induced smectic fluctuations of dynamical character, while the second one accounts for possible static fluctuations induced by any kind of pore wall irregularities, and thus by a QD field \cite{Bellini2001, *guegan:011707,*lefort:040701}. In absence of specific data for 12CB in the literature, values for $c$ and for the ratio $\frac{\xi_{\parallel}}{\xi_\bot}$ were assumed similar to those of 8CB \cite{guegan:011707,*lefort:040701}. We find that the broadening of the Bragg peak is entirely attributable to the static disorder term at all $T$s investigated. An equally good fit could be obtained by setting $A_{therm}=0$. This differs from 8CB in pSi, which exhibits pre-transitional thermal fluctuations reminiscent of the continuous character of the N-SmA transition in the bulk \cite{*guegan:011707,*lefort:040701, Lehenypre}. On the other hand, it is consistent with the fact that we do not observe any pre-transitional smectic diffuse scattering above 326~K. The temperature dependence of the smectic correlation length $\xi_{\parallel}$ extracted from the fitting procedure is shown in Fig.\ref{fig:1}c. 
For confined 12CB, $\xi_{\parallel}$ rises continuously to moderate values upon decreasing $T$. This is in striking contrast with bulk 12CB, where we observe a jump of $\xi$ at $T_{IA}^{bulk}$ to resolution limited values. This observation provides another strong evidence that the I-SmA transition of 12CB in anisotropic pSi actually loses its first order character unlike 10CB in isotropic random confinement, where a discontinuous variation of $\xi_{\parallel}$  was retained \cite{Bellini-10CB-JPCM-2003,*10CB-aerosil-PhysRevE.2004}.
\begin{figure}[h!]
\includegraphics[width=0.4\textwidth]{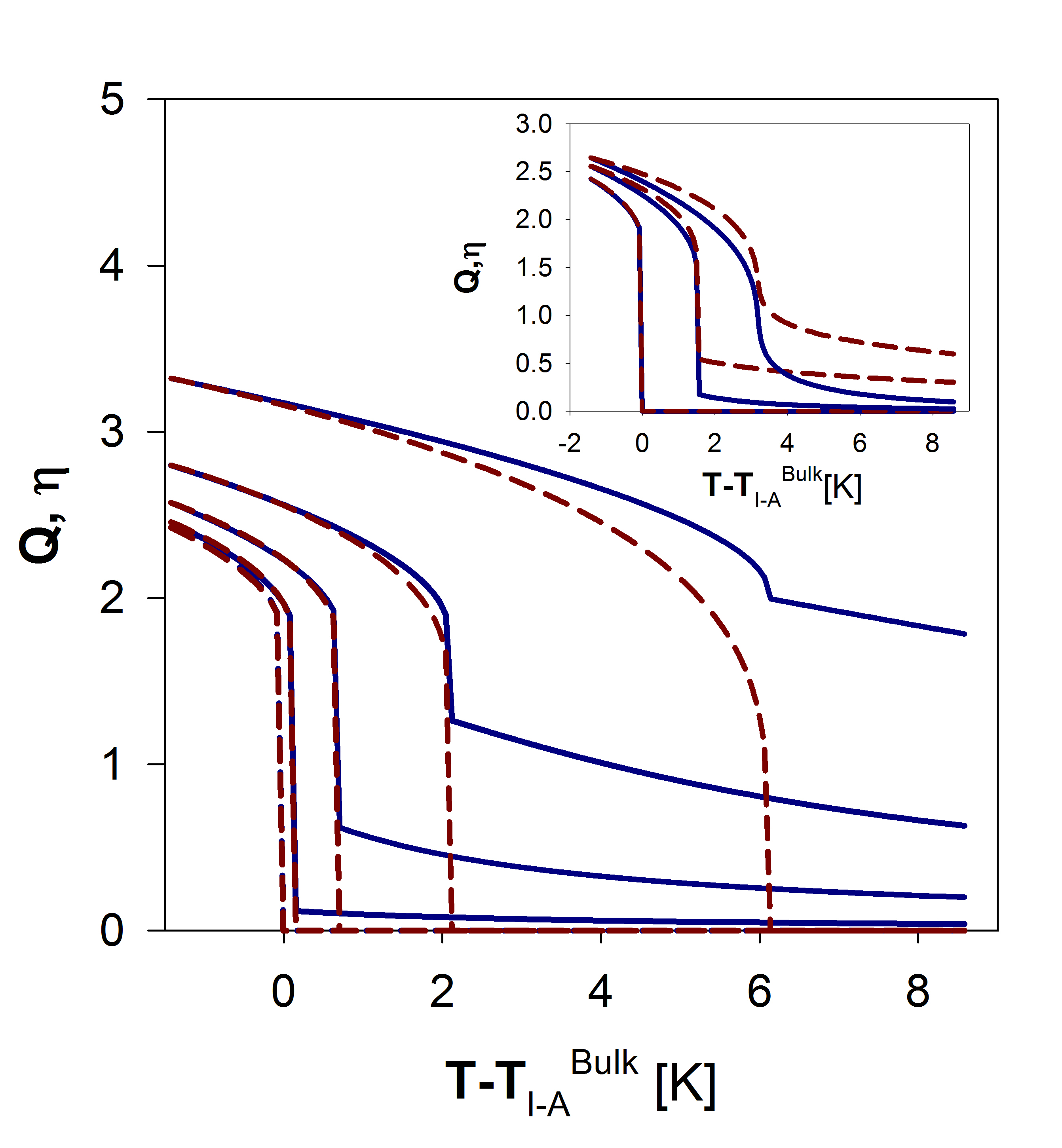} 
\caption{\label{fig:kklz} (Color online) Equilibrium solutions of the KKLZ model in the strong NS coupling limit ($D$=2000). All curves are normalized at low $T$, and so that $Q(T_{I-N})=$1 for a hypothetic system with $D=0$. Dashed lines : nematic OP ($Q$), solid lines : smectic OP ($\eta$). Main figure: solutions for $\sigma_n$ = 0, 1, 5, 15 and 50 and $\sigma_s$ = 0.  Inset: solution for $\sigma_n$ = 0 and $\sigma_s$ = 0, 0.002 and 0.005. }
\end{figure}
It is noteworthy that in case of a strong $Q$-$\eta$ coupling, the criticality of the I-SmA phase transition might result from the coupling between the nematic director and the unidirectional porous geometry, as in the case of the I-N transition of 8CB \cite{kityk:187801}, but also from the anisotropic QD effects. These two aspects are discussed in what follows. First we examine if this overall new phenomenology could be rationalized in the frame of a Landau-de~Gennes approach. The KKLZ model \cite{PhysRevE.54.1610,*PhysRevE.68.021705, *PhysRevE.70.051703, *PhysRevA.26.1610} applied to nCB (n$>$10) accounts for the $Q$-$\eta$ coupling, and also for the anisotropic coupling of the porous geometry with $Q$ ("field" effect) \cite{cordoyiannis:051703}. The reduced free energy then writes :
\begin{align}
\label{eqn:freenrj}
g = & t_n Q^2-2Q^3+Q^4-\sigma_n Q +\Delta t_n^{dis} Q^2 \\ \nonumber
& +\Omega \lbrace t_s \eta^2+\dfrac{\eta^4}{2}+\varepsilon \eta^6- \sigma_s \eta \rbrace \\ \nonumber
& -D Q \eta^2
\end{align}

where $t_n$ and $t_s$ are reduced temperatures, $\sigma_{n}$ and $\sigma_{s}$ represent the coupling terms of resp. nematic and smectic orderings with the unidirectional field imposed by the porous solid, and $D$ is the coupling strength between nematic and smectic OPs. While $t_n$ and $t_s$ only depend on intrinsic properties of the LC, $\sigma_{n}$ and $\sigma_{s}$ depend both on the LC and on the topology of the confining matrix \cite{cordoyiannis:051703, Kralj1995}. The reduced free energy $g$ is normalized with respect to nematic ordering, which implies a dimensionless prefactor $\Omega$ weighting the smectic contribution. $\Omega$ as well as $\varepsilon$ are material dependent. We select $\Omega\simeq79000$ to map experimental data of the family of cyanobiphenyls according to refs. \cite{*PhysRevA.26.1610, Kralj1995}. In the KKLZ model, disorder effects are simply taken into account by a renormalization (lowering) of the transition temperatures ($t_{n}\rightarrow t_{n}^{dis}=t_{n} + \Delta t_{n}^{dis}$) \cite{PhysRevE.68.021705}. Fig.\ref{fig:kklz} shows the equilibrium values of both nematic $Q$ and smectic $\eta$ OPs as computed from the KKLZ model with a $Q$-$\eta$ coupling parameter $D$ = 2000 (strong coupling limit), without disordering terms ($\Delta t_n^{dis}=0$). Clearly, both $\sigma_n$ and $\sigma_s$ field parameters induce an increase of the transition temperature. The opposite shift observed in the experiment can be qualitatively accounted for by a non-zero disorder term $\Delta t_{n}^{dis}$ in the free energy expansion (Eq. \ref{eqn:freenrj}) without changing the shape of the curves. 
Above the computed transition temperature, we see that the nematic field parameter $\sigma_n$ promotes paranematic ordering. This appears as a general feature of unidirectional confinement of LCs \cite{kityk:187801, PhysRevE.68.021705}. On the other hand, this orientational field contribution does not induce similar para\textit{smectic} order, even in this strong coupling regime. 
These two predictions qualitatively agree with experimental observations above $T_{IA}^{conf}$, provided the field parameter $\sigma_n$ is rather weak (between 1 and 1.5). These values are in agreement with the amplitude of field effects attributed to pSi through the analysis of the $I-N$ transition of confined 7CB and 8CB \cite{kityk:187801}. However, the model does not catch the essential characteristics of the transition itself. Even for very strong $\sigma_n$ values, it rather predicts a remaining first order character of the smectic ordering occuring from a paranematic or nematic high $T$-phase. Taking into account an additional smectic field term ($\sigma_s\neq$ 0, see inset of Fig. \ref{fig:kklz}) the system is driven accross a critical threshold, separating continous from discontinuous smectic ordering, but it also displays a strong parasmectic order at high $T$, in obvious contrast to the experimental observation (see Fig.\ref{fig:1}).

\begin{figure}
\includegraphics[width=0.4\textwidth]{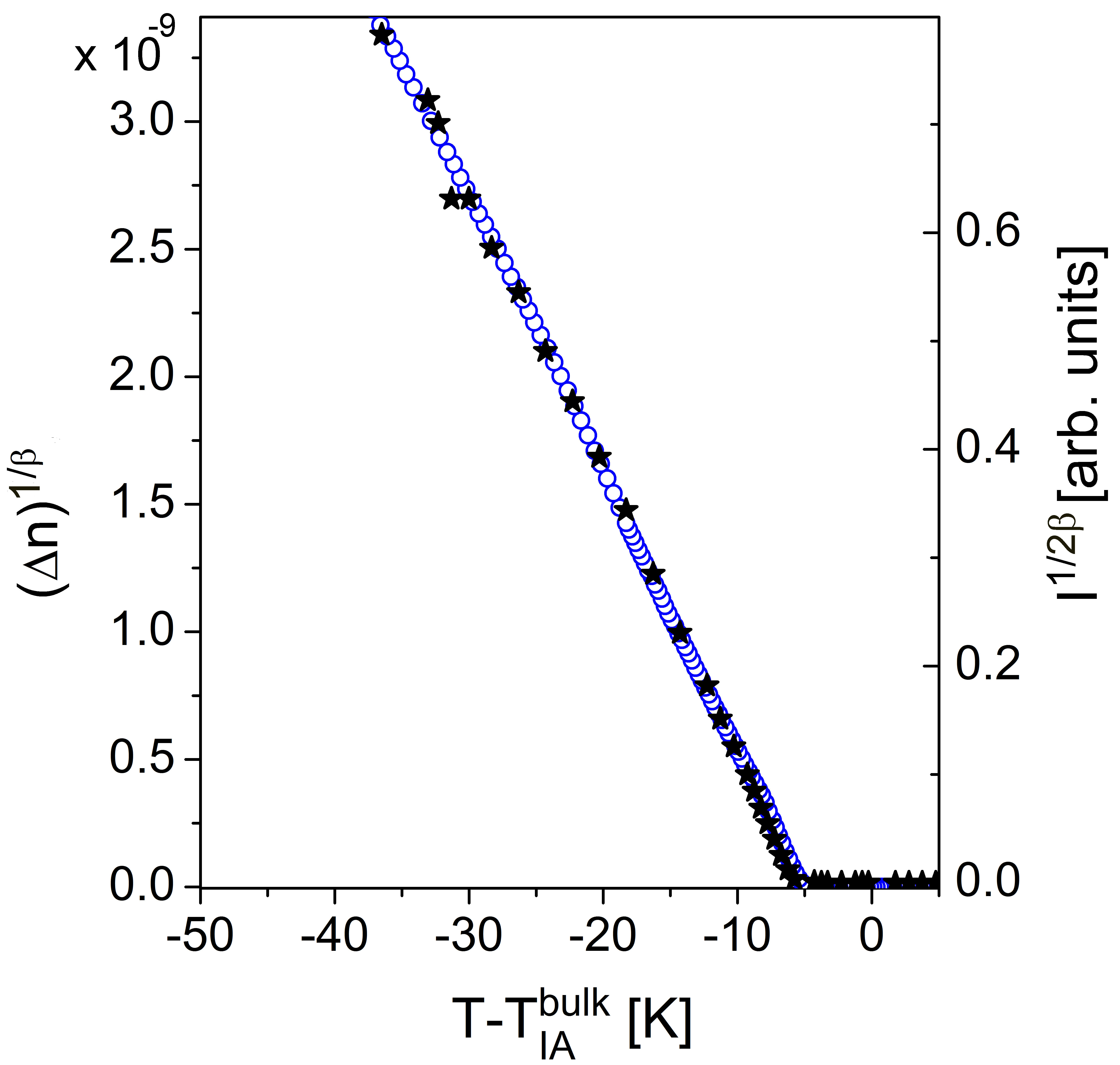} 
\caption{\label{fig:2} (Color online) Integrated intensity of the smectic peak to the power $\frac{1}{2\beta}$ and birefringence to the power $\frac{1}{\beta}$ with $\beta$ = 0.17. 
}
\end{figure}

These discrepancies suggest that random field effects play here a dominant role, which is supported by the observed downshift of $T_{IA}$ and the typical Lorentzian squared line shape of the diffraction pattern. As expected from Garland and Iannachione arguments \cite{Liang-8CB-aerosil-PRE-2007,*Iannacchine-8CB-PRE-2009,*Iannacchine-8CB-JPCB-2009}, the transition here has crossed a tricritical point under anisotropic QD, in a same way as if the $Q$-$\eta$ coupling would be turned off \cite{Gennes1976(2ndedition)}. The actual continuous character of the transition can be checked by inspecting the scalings of nematic and smectic OPs as a function of $T$. As shown in Fig.\ref{fig:2}, both OPs convincingly scale with the power law $Q,\eta\propto(T_c - T)^{\beta}$ with $T_c\approx326.4$ K. However, the corresponding best fit value of $\beta$ = 0.17$\pm$0.01 is significantly lower than the one expected in the frame of the undisturbed 3D~$XY$ model, and which was considered as the limit observed for nanoconfined 8CB under QD \cite{PhysRevE.49.2964,PhysRevB.60.206,*Iannacchine-8CB-PRE-2009,*Iannacchine-8CB-JPCB-2009}. It is rather comparable to temperature scalings observed for 8CB in aerogels or pSi \cite{Bellini2001,*guegan:011707}.  The observation of a low $\beta$ value for a continuous transition is in fact not unusual, and is observed in the case of dilute antiferromagnets or disordered ferroelectrics \cite{0295-5075-57-1-014, PhysRevB.74.144431}, which are prototypical realizations of the three dimensional random field Ising model (3DRFIM).


In conclusion, we report combined measurements of orientational and translational OPs of a LC confined in columnar pSi, characterized in the bulk by a strong first order nucleation-growth mechanism. In the high-$T$ phase, the confinement induces para\textit{nematic} but no para\textit{smectic} order. This fact can be interpreted as resulting from the unidirectional geometry of pSi, acting as an external field coupling to the nematic OP, as in the case of the $I-N$ transition of other cyanobiphenyls. In the case of 12CB, such field effects are not expected to be strong enough to modify the first order character of the $I-A$ transition. At $T_{IA}^{conf}$ however, no discontinuity of the OPs nor of the smectic correlation lengths is observed. Both $Q$ and $\eta$ grow according to the same power law. These observations rather indicate a critical-like behavior at low temperatures, which cannot be simply accounted for by available phenomenological considerations. Together with the downshift of $T_{IA}$ and the typical Lorentzian squared line shape of the diffraction pattern, they strongly support the prevalence of QD effects. 
Although up to now no available theoretical frame can account for the observed transition behaviour, our observation of a unique scaling law for both nematic and smectic OPs emphasizes that the phenomenology of LCs under anisotropic QD cannot be reduced to a simple tuning of the de Gennes' $Q$-$\eta$ coupling by the strength of the random field.

\bibliography{biblio_12CB_pSi}

\end{document}